\documentclass[smallextended]{svjour3}     
\pdfoutput=1
\smartqed  
\usepackage{graphicx}
 \usepackage{mathptmx}      
\usepackage[round]{natbib}
\usepackage{latexsym}
\usepackage{wasysym}
\usepackage{marvosym}

\usepackage{hyperref}               
\hypersetup{pdfauthor=Christoph Mordasini}
\hypersetup{backref=true, pagebackref=true, hyperindex=true, breaklinks=true,colorlinks=true,urlcolor=blue, linkcolor=blue,  citecolor=blue,pagecolor=red, bookmarks=true, bookmarksopen=true}

\newcommand{\bliss}{B_{\rm Liss}}
\newcommand{\mearth}{M_\oplus}

\newcommand{\rhill}{R_{\rm H}}

\newcommand{\mstar}{M_*}

\newcommand{\astart}{a_{\rm start}}

\newcommand{\miso}{M_{\rm iso}}

\newcommand{\aice}{a_{\rm ice}}
\newcommand{\beq}{\begin{equation}}
\newcommand{\eeq}{\end{equation}}

\begin{document}

\title{Theory of planet formation
}

\titlerunning{Theory of planet formation}        

\author{Christoph Mordasini        \and
        Hubert Klahr \and
        Yann Alibert \and
        Willy Benz \and Kai-Martin Dittkrist 
}

\authorrunning{Mordasini et al.} 

\institute{C. Mordasini \and H. Klahr \and K.-M. Dittkrist 
             \at
              Max Planck Institute for Astronomy, K\"onigstuhl 17, D-69117 Heidelberg,  Germany\\
              \email{mordasini@mpia.de, klahr@mpia.de}         \\
           \and
               Yann Alibert \and Willy Benz \at
              Physikalisches Institut, Universit\"at Bern, Sidlerstrasse 5, CH-3012 Bern, Switzerland\\
              \email{alibert@space.unibe.ch, benz@space.unibe.ch}         \\
}

\date{Received: date / Accepted: date}

\maketitle

\begin{abstract}
We review the current theoretical understanding how growth from micro-meter sized dust to massive giant planets occurs in disks around young stars. After introducing a number of observational constraints from the solar system, from observed protoplanetary disks, and from the extrasolar planets, we simplify the problem by dividing it into a number of discrete stages which are assumed to occur in a sequential way. In the first stage - the growth from dust to kilometer sized planetesimals - the aerodynamics of the bodies are of central importance. We discuss both a purely coagulative growth mode, as well as a gravoturbulent mode involving a gravitational instability of the dust. In the next stage, planetesimals grow to protoplanets of roughly 1000 km in size. Gravity is now the dominant force. The mass accretion can be strongly non-linear, leading to the detachment of a few big bodies from the remaining planetesimals. In the outer planetary system (outside a few AU), some of these bodies can become so massive that they eventually accrete a large gaseous envelope. This is the stage of giant planet formation, as understood within the core accretion-gas capture paradigm. We also discuss the direct gravitational collapse model where giant planets are thought to form directly via a gravitational fragmentation of the gas disk. In the inner system, protoplanets collide in the last stage - probably after the dispersal of the gaseous disk - in giant impacts until the separations between the remaining terrestrial planets become large enough to allow long term stability. We finish the review with some selected questions.   
 
\end{abstract}

\section{Introduction}\label{mordasinisec:intro}
In this article, we give an introductory overview of the theory of planet formation. As a short introduction, it can necessarily only provide a simplified description of the most important lines of reasoning of this theory, and many important aspects must remain unaddressed. For more comprehensive works, we refer the reader to \citet{lissauer1993a,papaloizouterquem2006,armitage2007}, and the book of  \citet{klahrbrandner2006}.

Our current understanding of planet formation is based on several centuries of studies of our own solar system, and additionally 15 years of detections of extrasolar planetary systems. This has lead to a general perception of planet formation as a process that can be described in a number of subsequent stages. It is clear that such a sequential picture is already a simplification, as the different stages will in reality at least partially overlap, because their duration is for example in general a function of the distance from the star. The different stages come along with a growth by at least thirteen orders of magnitude in spatial scale from the initially tiny dust grains to the giant planets, i.e. a huge dynamic range. The basic stages also give the structure of this review: In a first step, the gravitational collapse of a dense gas cloud forms a protostar with a surrounding protoplanetary disk consisting of gas and dust. This stage was addressed in the chapter by R. Alexander. In the second step which is discussed in section \ref{mordasinisec:dusttoplanetesimals}, the initially micrometer sized dust grows either via coagulation, or via an instability in the dust layer to form kilometer sized planetesimals. In a third step (section \ref{mordasinisec:planetesimalstoprotoplanets}), these planetesimals grow through two-body collisions to form protoplanets, with sizes of typically a few thousand kilometers. Some of these protoplanets might grow so large that they can accrete massive hydrogen/helium envelopes, and become giant planets (section \ref{mordasinisec:giantplanets}). Other remain too small for gas accretion to be effective. These protoplanets collide in the inner system after the dispersal of the disk to form terrestrial planets (section \ref{mordasinisec:terrestrialplanets}). The formation both of giant and terrestrial planets is influenced by orbital migration, i.e. a change in the semimajor axis of the protoplanets due to angular momentum exchange with the disk (section \ref{mordasinisec:migration}). In section \ref{mordasinisec:keyquestions} we discuss the two selected areas of active research, namely the place of origin of giant planets, and their luminosities.

\subsection{Observational constraints}\label{mordasinisec:obsconstraints}
The guidelines to understand the physics involved in the different stages of planet formation come from observational constraints derived from three different classes of astrophysical objects. The first is our own planetary system, i.e. the solar system. Studies of the solar system include remote observations of the sun, the planets and the minor bodies, laboratory analysis of meteorites, in-situ measurements by space probes, possibly including sample returns, as well as theoretical work and numerical modeling. No other planetary system can be studied in such detail as the solar system, and thus poses as detailed constraints on formation models. The solar system is therefore a necessary benchmark for all formation theories. On the other hand, the solar system alone is insufficient to fully capture the wealth of different physical mechanisms that are at work during planet formation, as shown by the fact that some extrasolar planets have properties which are very distinct from those of the solar system planets. 

The second class of astrophysical objects leading to important constraints on planetary formation are  protoplanetary disks. As planets are believed form in protoplanetary disks, the conditions in them are the initial and boundary conditions for the formation process. For several decades, nearby star formation regions have been studied to infer the various stages of disk formation and evolution, to derive the distributions of sizes, masses, radial structures of disks, to analyze the chemical processes occurring in them, and to determine  the lifetime of circumstellar disks. Regarding these lifetimes, photometric observations can measure the infrared excess caused by hot micrometer sized dust, and this excess radiation can be understood as a proxy for the presence of a primordial circumstellar disk. The observations from \citet{haischlada2001} or  \cite{hernandezhartmann2008} show that the fraction of stars having such an excess is a roughly linearly decreasing function of cluster age, disappearing after about 4-6 Myr. Giant planets must have formed at this moment. This represents a non-trivial constraint on classical giant planet formation models, as we shall see below.
 
The third class are finally the extrasolar planets, which can all be regarded as different examples of the final outcome of the formation process. Before 1995, it was commonly assumed that giant planets should only be found, similar to Jupiter, outside a few astronomical units, where more solids are available due to the jump in the abundance of planet building blocks at the ice condensation distance (see below), even though  planetary migration had been studied theoretically much earlier \citep{goldreichtremaine1980}. Then, in 1995 \citeauthor{mayorqueloz1995} discovered the first extrasolar planet around a solar type star, 51 Peg b, which is a Jovian mass planet which is orbiting the host star at a distance of just 1/20 the Earth-Sun distance, and this discovery was only a starting point to the currently nearly 500 detected extrasolar planets. Our knowledge about extrasolar planets today is still strongly affected by various observational detections biases. Nevertheless, one must build a planet formation theory in agreement with following observational results: 
\begin{itemize}
\item Extrasolar planets orbit a significant fractions of solar like stars. About  10 \%  have giant planets within a few astronomical units, and roughly 30\%  have low mass planets, as indicated by recent results of \citet{mayorudry2009}. 
\item There is a huge diversity of extrasolar planets  both in mass and distance from the parent star. The observed distribution of extrasolar planets in the mass-distance diagram (see the chapter by F. Pont) has become a representation of similar importance for formation theories as the Hertzsprung Russell diagram for stellar astrophysics \citep{idalin2004}.  There are very massive ÒSuper JupiterÓ planets (with masses exceeding ten times the mass of Jupiter),  many  ÒHot JupitersÓ like 51 Peg b, Hot Neptunes and Super-Earth planets, or planets on very eccentric orbits. All these findings were not necessarily expected from the shape of our own planetary system. 
\item Heavy elements as measured by the metallicity [Fe/H] of the host star play an important role in the formation of at  least the giant planets \citep[e.g.][]{santosisraelian2004,fischervalenti2005}.
\end{itemize}
Especially the fact that many extrasolar planets were found exactly where one did not expect to find them pointed towards a serious gap in the understanding of planet formation derived from the solar system alone, so that the mentioned orbital migration which we will address in section \ref{mordasinisec:migration} is nowadays regarded as an integral component of modern planet formation theory.

The high number of extrasolar planets also gives us the possibility to look in a new way at all these discoveries: To see them no more just as single objects (although some are also individually very interesting), but to see them as a population. Thus, we can look at distributions of extrasolar planet masses, semimajor axes, host star metallicities and so on, as well as all kinds of correlations between them. A theoretical study of these statistical properties of the exoplanet population is done best by the method of planetary population synthesis \citep{idalin2004,idalin2004a,mordasinialibert2009a,mordasinialibert2009b}, a technique which was also used to obtain some of the results discussed here (section \ref{mordasinisec:migration} and \ref{mordasinisec:keyquestions}).

\section{From dust to planetesimals}\label{mordasinisec:dusttoplanetesimals}
The first stage of planetary growth starts with roughly micro-meter sized dust grains, similar as those found in the interstellar medium. These tiny objects are well coupled  to the motion of the gas in the protoplanetary disk via gas drag. With increasing mass, gravity becomes important also, and the particles decouple from the pure gas motion. This stage involves the growth of the dust grains via coagulation (sticking), their sedimentation towards the disk midplane, and their radial drift towards the star.  In this context it is important to see that the gas and the solid particles move around the star at a slightly different orbital speed. The reason for this is that the gas is partially pressure supported (both the centrifugal force and the gas pressure counteract the gravity), and therefore moves slightly slower (sub-Keplerian) around the star. The resulting gas drag in turn causes the drift of the solid particles towards the star.

\subsection{Classical coagulation}\label{mordasinisect:classicalcoagulation}
\begin{figure}\sidecaption
\resizebox{0.6\hsize}{!}{\includegraphics*{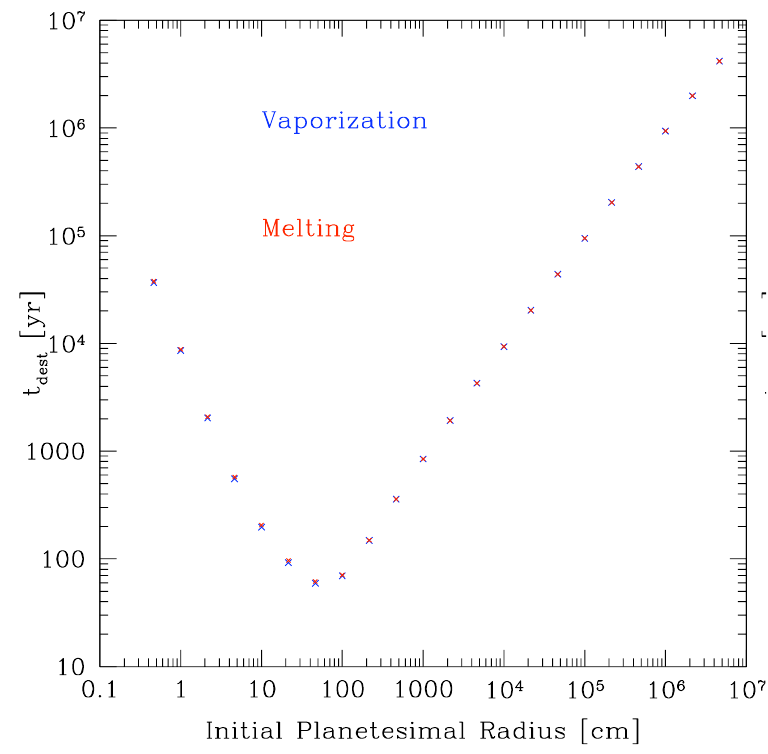}}
\caption{The time until destruction by spiraling in close to the host star for bodies of different initial sizes, starting initially at 1 AU.  The times were obtained by direct orbit integration under the influence of gas drag and gravity, and mass loss through thermal ablation occurring either via melting or vaporization. From \citet{mordasini2008}.}\label{mordasinifig:drift}
\end{figure}

In the picture of classical coagulation, bodies grow all the way to kilometer size by two body collisions. While growth from dust grains to roughly meter sized bodies can be reasonably well  modeled with classical coagulation simulations as for example in \citet{brauerdullemond2008}, two significant problems arise at the so called ``meter barrier'':
\begin{itemize}
\item For typical disk properties, objects which are roughly meter sized drift extremely quickly towards the central star where they are destroyed by the high temperatures. The drift timescale at this size becomes in particular shorter than the timescale for further growth \citep{klahrbodenheimer2006}, so that growth effectively stops. The very fast drift of roughly meter sized bodies is illustrated in Fig. \ref{mordasinifig:drift}. Note that the time until destruction by thermal ablation close the star for roughly meter sized bodies starting at 1 AU is less than just 100 years.\item The second problem arises from the fact that meter sized boulders do not stick well together, but rather shatter at the typical collision speeds which arise from the turbulent motion of the disk gas, and the differential radial drift  motion. 
\end{itemize}

\subsection{Planetesimal formation by self gravity}\label{mordasinisect:planetesimalselfgravity}
First ideas how to bypass the critical meter size were put forward already a long time ago, and invoke the instability of the dust layer to its own gravity. Here, one can quickly jump from small sub-meter objects to full blown planetesimals.  In the classical model of \citet{goldreichward1973}, dust settles into a thin layer in the disk midplane. If the concentration of the dust becomes sufficiently high (see section \ref{mordasinisec:directcollapse} for a similar condition for gas), the dust becomes unstable to its own gravity and collapses to from planetesimals directly. The turbulent speed of the grains  however must  be very low to reach the necessary concentration. This condition is difficult to meet, as the vertical velocity shear between the dust disk rotation at the Keplerian frequency and the dust poor gas above and below the midplane rotating slightly sub-Keplerian causes the development of  Kelvin-Helmholtz instabilities. The resulting turbulence is sufficiently strong to decrease the particle concentration below the threshold necessary for the gravitational collapse. This is why self-gravity of the dust, and gas turbulence, either due to the Kelvin-Helmholtz mechanism, or due to the magneto-rotational instability \citep{balbushawley1998}, were for a long time thought to be mutually exclusive.

In the recent years however, significant progress has been made in the direction of planetesimal formation by self gravity \citep{johansenklahr2006,cuzzihogan2008}, and it was in particular understood that turbulence can actually aid the formation of planetesimals, rather than hindering it. The reason is that turbulence can locally lead to severe over-densities of the solid particle concentration by a factor as high as 80 compared to the normal dust to gas ratio  on large scales of the turbulence \citep{johansenklahr2006} or by $\sim10^{3}$ on small scales \citep{cuzzihogan2008}. Further concentration can occur thanks to the streaming instability \citep{youdingoodman2005}, which all together can lead to the formation of gravitationally bound clusters with already impressive masses, comparable to dwarf planets \citep{johansenoishi2007}, on a timescale much shorter than the drift timescale.  

This so called gravoturbulent planetesimal formation should leave imprints visible in the solar system distinguishing it from the classical pure coagulation mechanism.  The recent work of \citet{morbidellibottke2009} finds that the observed size frequency distribution (SFD) in the asteroid belt cannot be reproduced with planetesimals that grow (and fragment) starting at a small size. It rather seems that planetesimals need to get born big in order to satisfy this observational constraint, with initial sizes between already 100 and 1000 km. 

On the other hand it should  be noted that too large initial sizes for the majority of the planetesimals might not be desirable either, as this could slow down the formation of giant planet cores, at least if the planetesimal accretion occurs through a mechanism similar as described in \citet{pollackhubickyj1996}.  This is due to the higher random velocities of massive planetesimals, and the less effective capture of larger bodies by the protoplanetary gaseous envelope.

\section{From planetesimals to protoplanets}\label{mordasinisec:planetesimalstoprotoplanets}
At the size of planetesimals (kilometers), gravity is the clearly dominant force, even though the gas drag still plays a role. The growth stage from planetesimals to protoplanets (with radii of order a thousand kilometers, corresponding to the isolation mass in the inner system, seen \citet{goldreichlithwick2004}, and below) however remains challenging to study because of the following reasons:
\begin{itemize}
\item The  initial conditions are poorly known, as the formation mechanism and thus the size distribution of the  first planetesimals is not yet clear as we have seen in the previous section.
\item One has to follow a very large number of planetesimals, thus no direct N-body integrations are possible with today's computational capabilities. For example  a planet of a mass of about ten Earth masses consists of more than $10^{8}$ planetesimals with a radius of 30 km.
\item The time that has to be simulated is long, typically several million years, equivalent to the same number of dynamical timescales (at 1 AU). 
\item The growth process is highly non-linear and involves complex feed-back mechanisms as the growing bodies play an increasing role in the dynamics of the system.
\item The physics describing the collisions which are ultimately needed for growth are non-trivial and include e.g. shock waves, multi-phase fluids and fracturing.
\end{itemize}
This growth stage has therefore been modelled with different methods, each having different abilities to address the listed issues. In this article we will only address the most basic approach. Other, more complex methods which yield a more realistic description of this stage are  statistical methods \citep[e.g.][]{inabatanaka2001} or Monte Carlo methods \citep[e.g.][]{ormeldullemond2010}.

\subsection{Rate equations}\label{mordasinisec:rateequations}
In this  approach one describes the growth of a body in the form of a rate equation which directly gives the mass growth rate $dM/dt$ of a body as a function of several quantities.  Usually it is assumed that one big body (the protoplanet) collisionally grows from the accretion of much smaller background planetesimal, see e.g. \citet{goldreichlithwick2004}. These background planetesimals are characterized by a size (or a size distribution), a surface density and a dynamical state (eccentricity and inclination). 

The growth rate is then described with a \citet{safronov1969} type equation
\beq
\frac{dM}{dt}=\pi R^2 \Omega\Sigma_{\rm p}  F_{\rm G}
\eeq
where $\Omega$ is the Keplerian frequency of the big body at an orbital distance $a$ around the star of mass $\mstar$, $\Omega=\sqrt{G \mstar/a^{3}}$, $\Sigma_{p}$ is the surface density of the field planetesimals, $R$ is the radius of the big body, and $F_{\rm G}$ is the gravitational focussing factor. It reflects the fact that due to gravity, the effective collisional cross section of the body is larger than the purely geometrical one, $\pi R^{2}$, as the trajectories get bended towards the big body.

The focusing factor is the key parameter in this equation, as it gives raise to different growth regimes like runaway, oligarchic or orderly growth, which come with very different growth rates \citep{rafikov2003a}. Its value is dependent on the random velocity  of the small bodies $v_{\rm ran}$ relative to the local circular motion. It scales with their eccentricity and inclination. The small bodies are affected by the encounters mutually between them, and with the big body (which increase the random velocity) and the damping influence by the gas (which decreases it).  

In the simplest approximation, where one neglects the influence of the sun, $F_{\rm G}$ is given as $1+v_{\rm esc}^{2}/v_{\rm ran}^{2}$, where  $v_{\rm esc}=\sqrt{2 G M/R}$ is the escape velocity from the big body. One notes that if the planetesimals have small random velocities in comparison with $v_{\rm esc}$, then $F_{\rm G}$ is large and fast accretion occurs. The mass accretion rate is then proportional to $R^{4}$, i.e. strongly nonlinear. This is the case in the so called runaway growth regime, where massive bodies grow quicker than small ones, so that these runaway bodies detach from the remaining population of small planetesimals \citep{weidenschillingspaute1997}.  With increasing mass, these big bodies however start to increase the random velocities of the small ones, so that the growth becomes slower, and the mode changes to the so called oligarchic mode, where the big bodies grow in lockstep \citep{idamakino1993}. 

\begin{figure*}
\begin{minipage}{0.5\textwidth}
	      \centering
       \includegraphics[width=\textwidth]{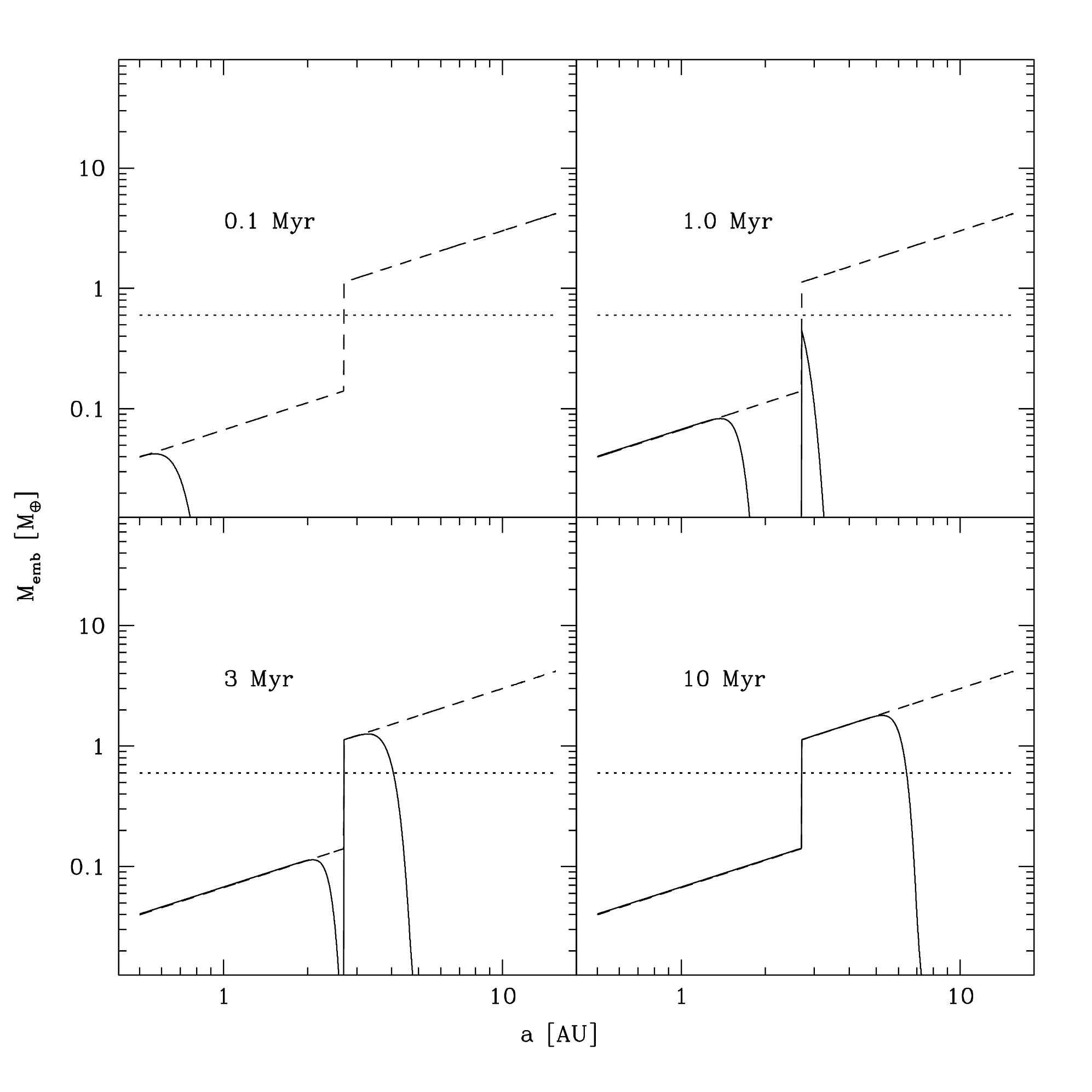}
     \end{minipage}\hfill
     \begin{minipage}{0.5\textwidth}
      \centering
       \includegraphics[width=\textwidth]{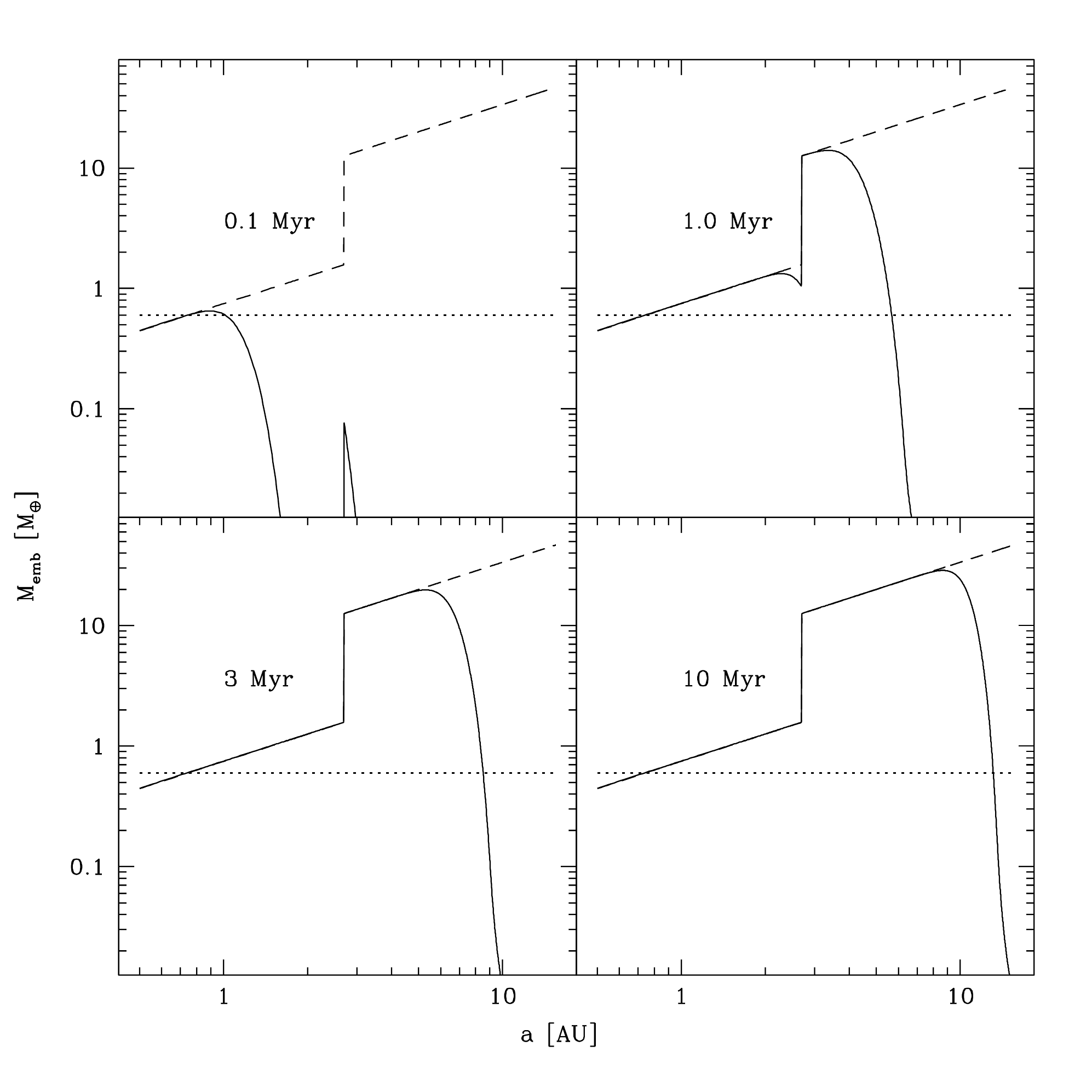}
     \end{minipage}
     \caption{Snapshots of  protoplanetary masses (solid line) as a function
     of semimajor axis at four moments in time for two different initial solid
     surface densities.  The dashed line is the isolation mass. The initial solid
     surface density at 1 AU is 7 g/cm$^2$  in the left panel and 35
     g/cm$^2$ in the right panel. From \citet{mordasinialibert2009a}. }\label{mordasinifig:rateequation}
 \end{figure*}

As the planet grows in mass, the surface density of planetesimals must decrease correspondingly  \citep[e.g.][]{thommesduncan2003}. As the growing protoplanet  is itself embedded in the gravitational field of the sun, one finds by studying the restricted three body problem that a growing body can only accrete planetesimals which are within its gravitational reach (in its feeding zone), i.e. in an annulus around the body which has a half width which is a multiple $\bliss\approx 4$ of the Hill sphere radius $\rhill$, 
\beq\label{mordasinieq:rhill}
\rhill=\left(\frac{M}{3 \mstar}\right)^{1/3}a
\eeq
 of the protoplanet.  Without radial excursions (migration), a protoplanet can therefore grow in situ only up to the so called isolation mass \citep{lissauer1993a}
  \beq\label{mordasinieq:miso}
\miso=  \frac{(4 \pi \bliss a^2 \Sigma_{\rm p})^{3/2}}{(3 \mstar)^{1/2}}.
\eeq 
Figure \ref{mordasinifig:rateequation} \citep[taken from][]{mordasinialibert2009a} shows snapshots of the  protoplanet mass as a function of semimajor axis for four different moments in time, obtained from numerically integrating a similar rate equation. The left panels shows a disk with a planetesimal surface density as in the minimal mass solar nebula (MMSN\footnote{In the MMSN model \citep{weidenschilling1977,hayashi1981} it is assumed that the planets formed at their current positions, and that the solids found in all planets of the solar system today correspond to the amount of solids available also at the time of formation. The total disk mass is then found by adding gas in the same proportion as observed in the sun.}), while in the right panel the surface density is five times as high. 

One sees from the plot that growth proceeds faster at smaller distances, but stops also at smaller masses (the isolation mass increases $\propto a^{3/4}$ for the radial dependence of  $\Sigma_{\rm p}(a)\propto a^{-3/2}$ used here). No giant planet can form in situ (cf. section \ref{mordasinisec:coreaccretion}). Fairly rapid growth to relatively large masses occurs just outside the ice condensation distance (the iceline $\aice$) which is here fixed to 2.7 AU.  In a disk with a higher solid surface density (right panel), protoplanets grow more rapidly, and also to higher masses, of order 10 $\mearth$ outside the iceline. 

The outcome of this growth stage is therefore in the inner planetary system a high number of  small (masses of 0.01 to 0.1 Earth masses $\mearth$) protoplanets (oligarchs). During the presence of the gas disk, growth is stalled at such masses, as gas damping hinders the development of high eccentricities which would be necessary for mutual collision between these bodies \citep{idalin2004}.  In the outer planetary system (beyond the iceline), a few 1 to 10 $\mearth$ protoplanets form. If such oligarchs reach during the disk lifetime a sufficiently high mass ($M\sim10\mearth$), they have the potential to trigger rapid gas accretion, as we will see below. 


\section{From protoplanets to giant planets}\label{mordasinisec:giantplanets}
For the formation of the most massive planets, the gaseous giants, two competing theories exist. The most widely accepted theory is the so called core accretion - gas capture model. We will however first discuss the direct gravitational collapse model.

\subsection{Direct gravitational collapse model}\label{mordasinisec:directcollapse}
In this model, giant planets are thought to form directly from the collapse of a part of the gaseous protoplanetary disk into a gravitationally bound clump. As we will see below, this requires fairly massive disks, so that it is thought that this mechanism should occur early in the disk evolution.  For the mechanism to work, two requirements must be fulfilled:

First, the self-gravity of the disk (measured at one vertical scale height) must be important compared to the gravity exerted by the star \citep{dangelodurisen2010}. The linear stability analysis of \citet{toomre1981} shows that a disk is unstable for the growth of axisymmetric radial rings if the Toomre parameter $Q$
\beq
Q=\frac{c_{\rm s} \kappa}{\pi G \Sigma}<1,
\eeq
where  $c_{\rm s}$ is the sound speed, $\kappa$ the epicyclic frequency (equal to Keplerian frequency in a strictly Keplerian disk), and $\Sigma$ is the gas surface density. From this equation we see that disks become unstable when they are cold ($c_{\rm}\propto T^{1/2}$), and massive (high $\Sigma$). 
Hydrodynamical simulations show that disks become unstable to non-axisymmetric perturbations (spiral waves) already at slightly higher $Q$ values of about $Q_{\rm crit}=1.4$ to 2 \citep{mayerquinn2004}. At the orbital distance of Jupiter, one finds that in order to become unstable, the surface density of gas must be about 10 times larger than in the MMSN.

Second, in order to allow fragmentation into bound clumps, the timescale $\tau_{\rm cool}$ on which a gas parcel in the disk cools and thus contracts must be short compared to the shearing timescale, on which the clump would be disrupted otherwise, which is equal the orbital timescale $\tau_{\rm orb}$
\beq
\frac{\tau_{\rm cool}}{\tau_{\rm orb}}<\xi_{\rm frag},
\eeq
where $\xi_{\rm frag}$ is of order unity \citep{gammie2001}. If this conditions is not fulfilled, only spiral waves develop leading to a gravoturbulent disk. The spiral waves transport angular momentum outwards, and therefore let matter spiral inwards to the star. This process liberates gravitational binding energy, which increases the temperature, and reduces the gas surface temperature, so that the disk evolves to a steady state of marginal instability only, without fragmentation.   

From the timescale arguments we see that the correct treatment of the disk thermodynamics is of central importance to understand whether the direct collapse model can operate. Radiation hydrodynamic simulations  give a confused picture, where different groups find for similar initial conditions both fragmentation \citep{boss2007} and no fragmentation \citep{caipickett2010}. This illustrates that the question whether bound clumps can form is still being debated. It is usually thought that it is unlikely that disk instability as a formation mechanism can work inside several tens of AU. This is due the fact, that the two necessary conditions discussed above lead to the following dilemma, as first noted by \citet{rafikov2005}: A disk that is massive enough to be gravitationally unstable is at the same time too massive to cool quickly enough to fragment, at least inside say 40-100 AU. Outside such a distance, the situation might be different as the orbital timescales become long \citep{boley2009}.  

\begin{figure}\sidecaption
       \includegraphics[width=0.6\textwidth,height=6cm]{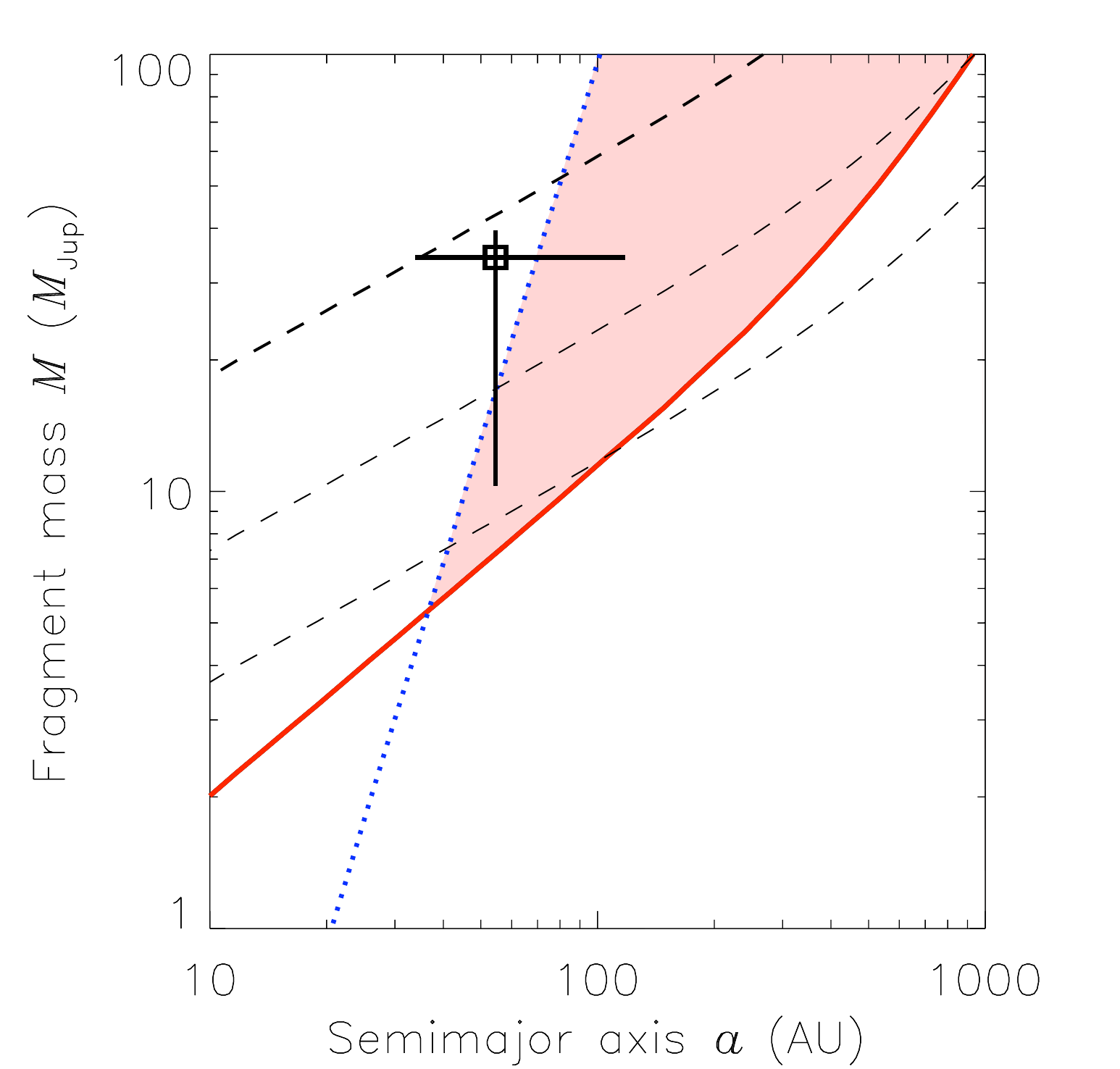}
     \caption{Direct gravitational collapse model for  GJ 758B. The black open square shows the position and mass, and vertical and horizontal black lines the uncertainties on the properties of this object. The red solid line is the minimal gas surface density (converted into a fragment mass) necessary for gravitational instability. The blue dotted line is the maximal  gas surface density allowing sufficiently fast cooling. The red area is therefore the domain where the gravitational collapse can operate. Dashed lines show the achievable fragment masses for disk masses of 0.1, 0.2, and 0.5 $\mstar$ (bottom to top). From Klahr et al. in prep. Reproduced with permission from the author.}\label{mordasinifig:directcollapse}
 \end{figure}

In figure \ref{mordasinifig:directcollapse} (Klahr et al. in prep.) it is studied whether GJ 758B which is a roughly 30 Jupiter mass brown dwarf at a semimajor axis of about 55 AU \citep{thalmanncarson2009} around a solar like star could have  formed by direct gravitational collapse. 

This study is based on the behavior of $Q$ and $\xi_{\rm frag}$ in vertical 1D disk models with realistic radiative cooling, taking into account the influence of the stellar mass, the stellar luminosity and the metallicity. Gas surface densities $\Sigma$ have been converted into planetary masses via the most unstable wavelength, c.f. \citet{dangelodurisen2010}. The red area gives the parameter space where the disk is both massive enough to be gravitationally unstable as indicated by the red solid line, but also cools quickly enough (blue dotted line). GJ 758B falls, within its error bars, into the allowed domain. Note that inside a distance of about 40 AU, planet formation through direct collapse can not occur, because in this part, the required minimum $\Sigma$ for instability is larger than the maximal $\Sigma$ allowed for a fast enough cooling.

In contrast to the core accretion model, giant planet formation is extremely fast  in the direct collapse model, as it occurs on a dynamical timescale.

\subsection{Core accretion model}\label{mordasinisec:coreaccretion}
The basic setup for the core accretion model is to follow the concurrent growth of a initially small solid core consisting of ices and rocks, and a surrounding gaseous envelope, embedded in the protoplanetary disk. This concept has been studied for over thirty years \citep{perricameron1974,mizunonakazawa1978,bodenheimerpollack1986}. Within the core accretion paradigm, giant planet formation happens as a two step process: first a solid core with a critical mass (of order 10 $\mearth$) must form, then the rapid accretion of a massive gaseous envelope sets in. 

The growth of the solid core by the accretion of planetesimals is modeled in the same way as described in section \ref{mordasinisec:planetesimalstoprotoplanets}. The growth of the gaseous envelope is described by one dimensional hydrostatic structure equations (similar to those  for stars), except that the energy release by nuclear fusion is replaced by the heating by impacting planetesimals $\epsilon$, which are the main energy source during the early phases. The other equations are the equation of mass conservation, of hydrostatic equilibrium, of energy conservation and of energy transfer:
\begin{eqnarray}
\frac{dr}{dm}&=&\frac{1}{4 \pi \rho r^{2}}\\
\frac{dP}{dm}&=&-\frac{G(m+M_{\rm core})}{4 \pi r^{4}}\\
\frac{dL}{dm}&=&\epsilon-T\frac{\partial S}{\partial t}\\
\frac{dT}{dP}&=&\nabla_{\rm ad} \mathrm{\ or\ } \nabla_{\rm rad}
\end{eqnarray}
In these equations, $r$ is the radius as measured from the planetary center, $m$ the gas mass inside r, $\rho, P, T, S$ the gas density, pressure, temperature and entropy, $t$ the time, and the temperature gradient can be given either by the radiative or the adiabatic gradient, whichever is shallower. To solve these equations one needs boundary conditions \citep{bodenheimerhubickyj2000,papaloizounelson2005}. We have to distinguish two regimes: 

At low masses, the envelope of the protoplanet is attached continuously to the background nebula, and the conditions at the surface of the planet  are just the pressure and temperature in the surrounding disk. The radius of the planet  is given in this regime approximately by the Hill sphere radius. The gas accretion rate is given by the ability of the envelope to radiate away energy so that it can contract, so that in turn new gas can stream in. 

When the gas accretion obtained in this way becomes too high in comparison with the externally possible gas supply, the planet enters the second phase and contracts to a radius which is much smaller than the Hill sphere radius. This is the detached regime of high mass, runaway (or post-runaway) planets. The planet now adjusts its radius to the boundary conditions that are given by an accretion shock on the surface for matter falling onto the planet from the Hill sphere radius, or probably more realistically, by conditions appropriate for the interface to a circumplanetary disk. In this phase, the gas accretion rate is no more controlled by the planetary structure itself, but by how much gas is supplied by the disk and can pass the gap formed by the planet in the protoplanetary disk \citep{lubowseibert1999}.

\citet{pollackhubickyj1996} have implemented the mentioned equations in a model that we may call the baseline formation model.  They assumed a constant pressure and temperature in the surrounding disk, and a strictly fixed embryo position, i.e. no migration.  

\begin{figure*}
\begin{minipage}{0.5\textwidth}
	      \centering
       \includegraphics[width=\textwidth]{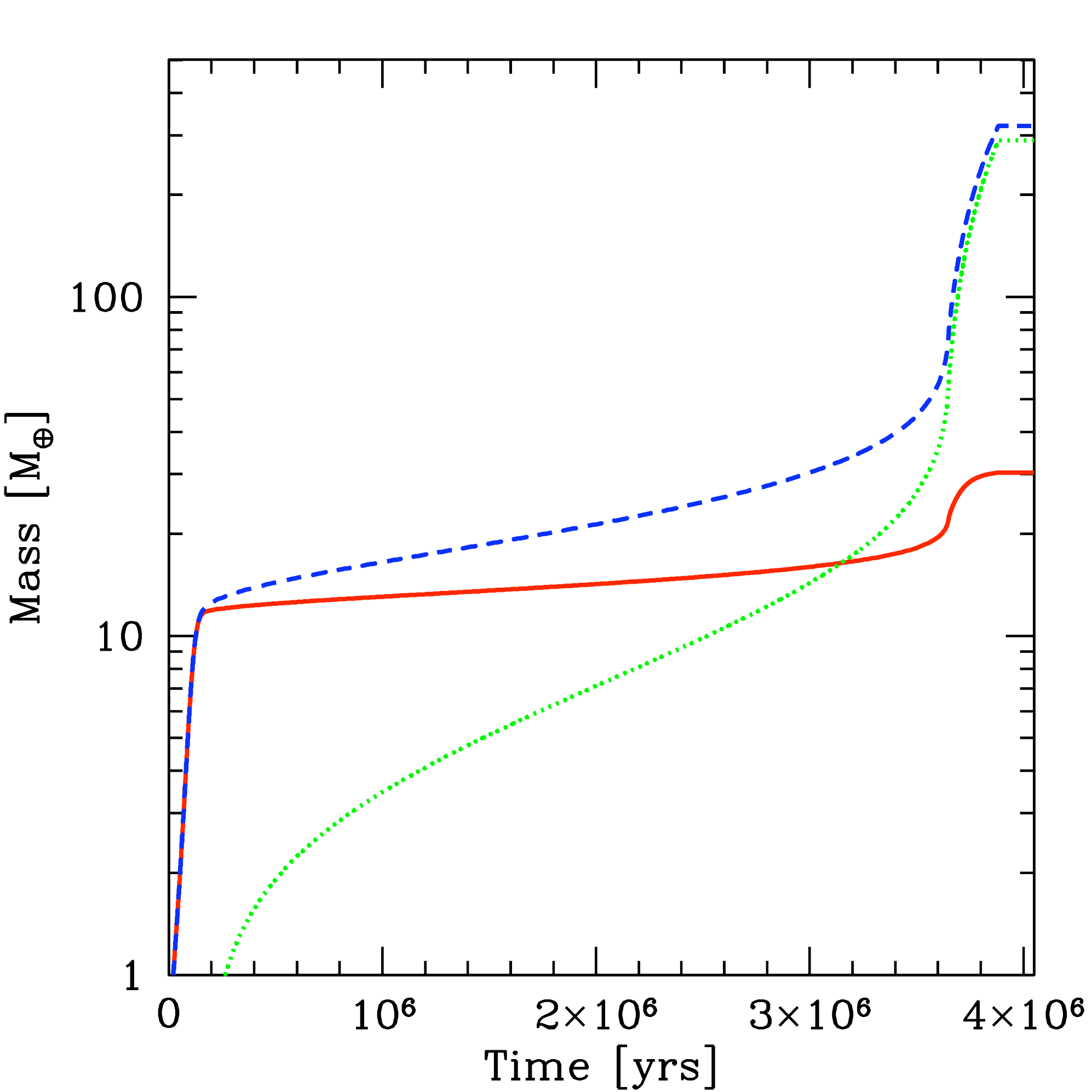}
     \end{minipage}\hfill
     \begin{minipage}{0.5\textwidth}
      \centering
       \includegraphics[width=\textwidth]{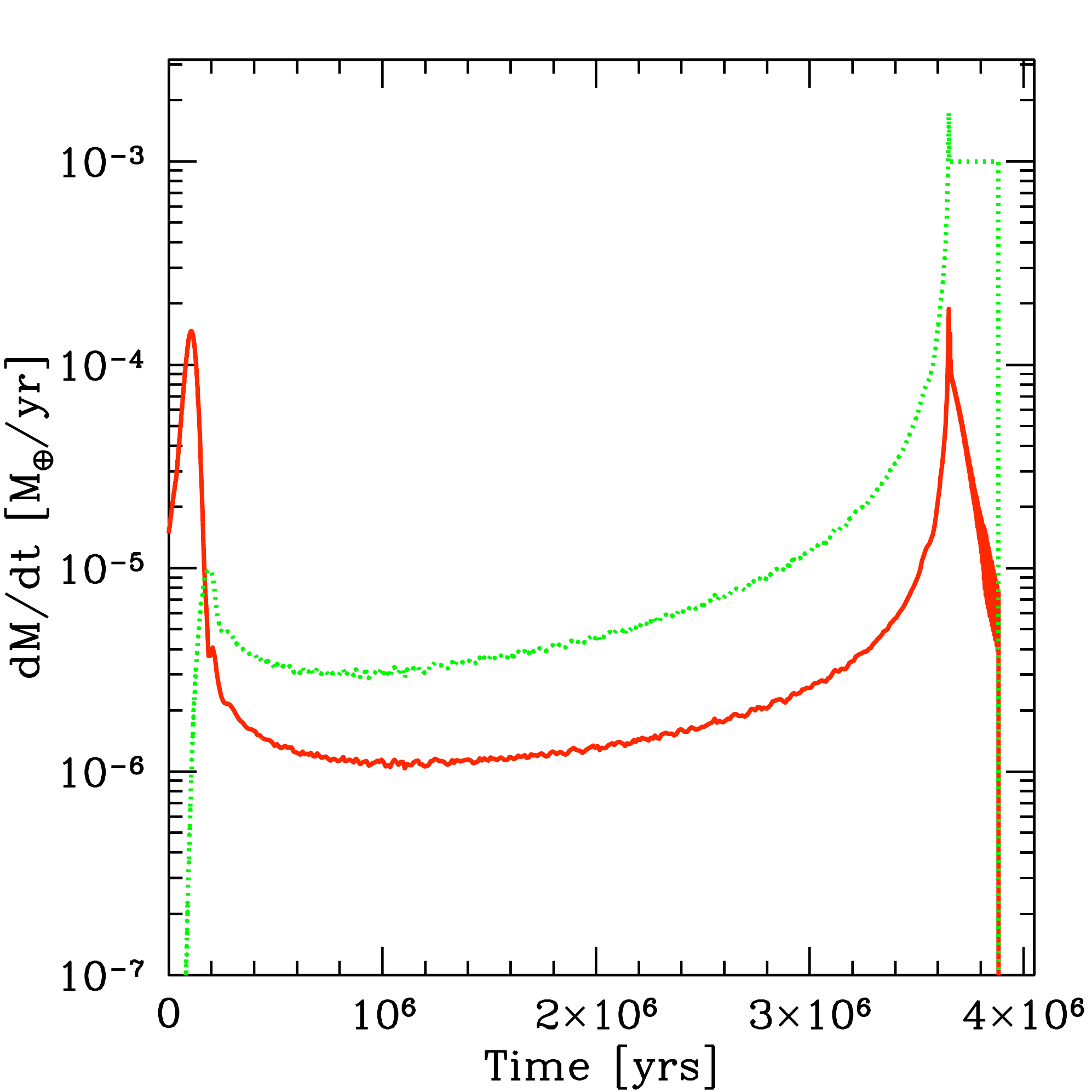}
     \end{minipage}
     \begin{minipage}{0.5\textwidth}
	      \centering
       \includegraphics[width=\textwidth]{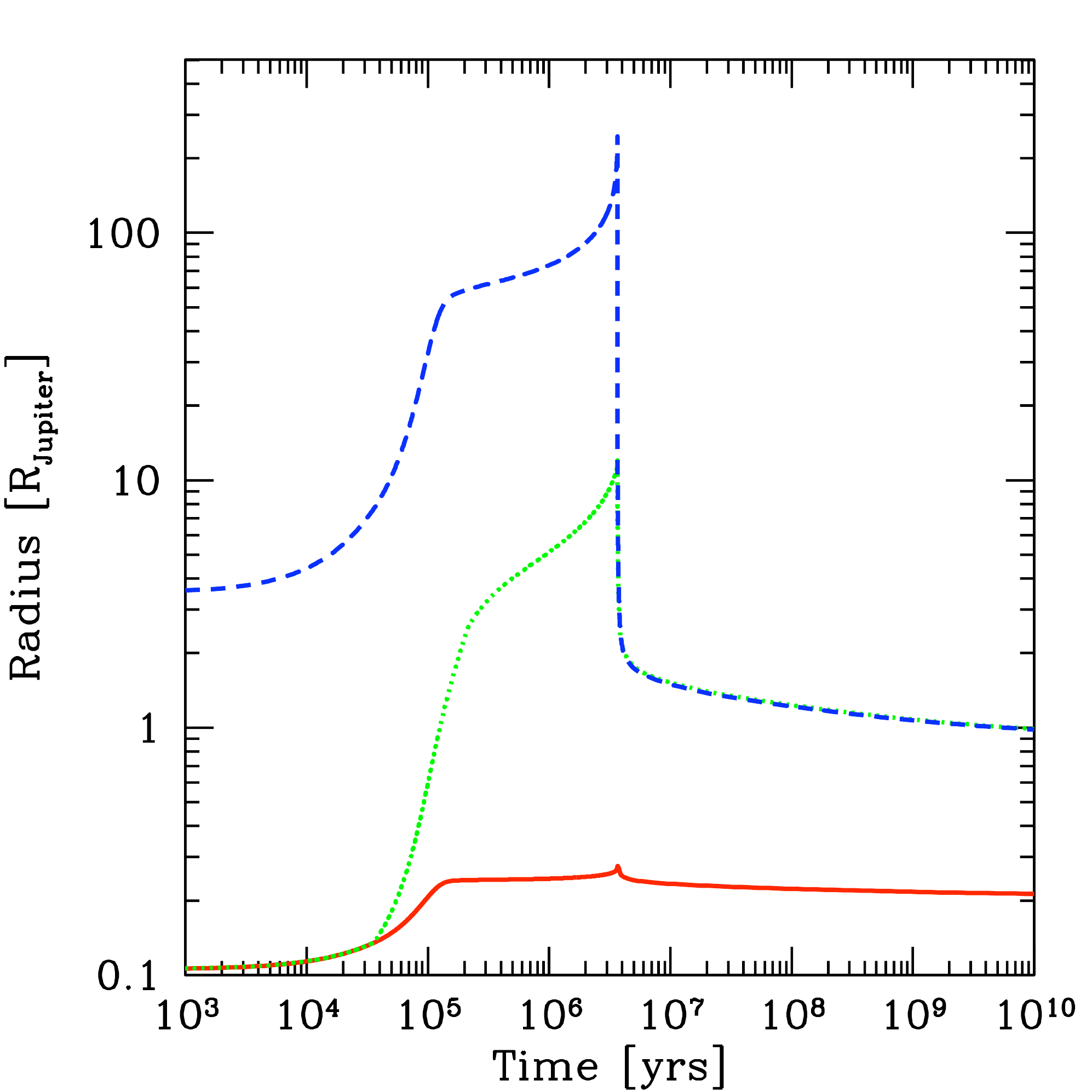}
     \end{minipage}\hfill
     \begin{minipage}{0.5\textwidth}
      \centering
       \includegraphics[width=\textwidth]{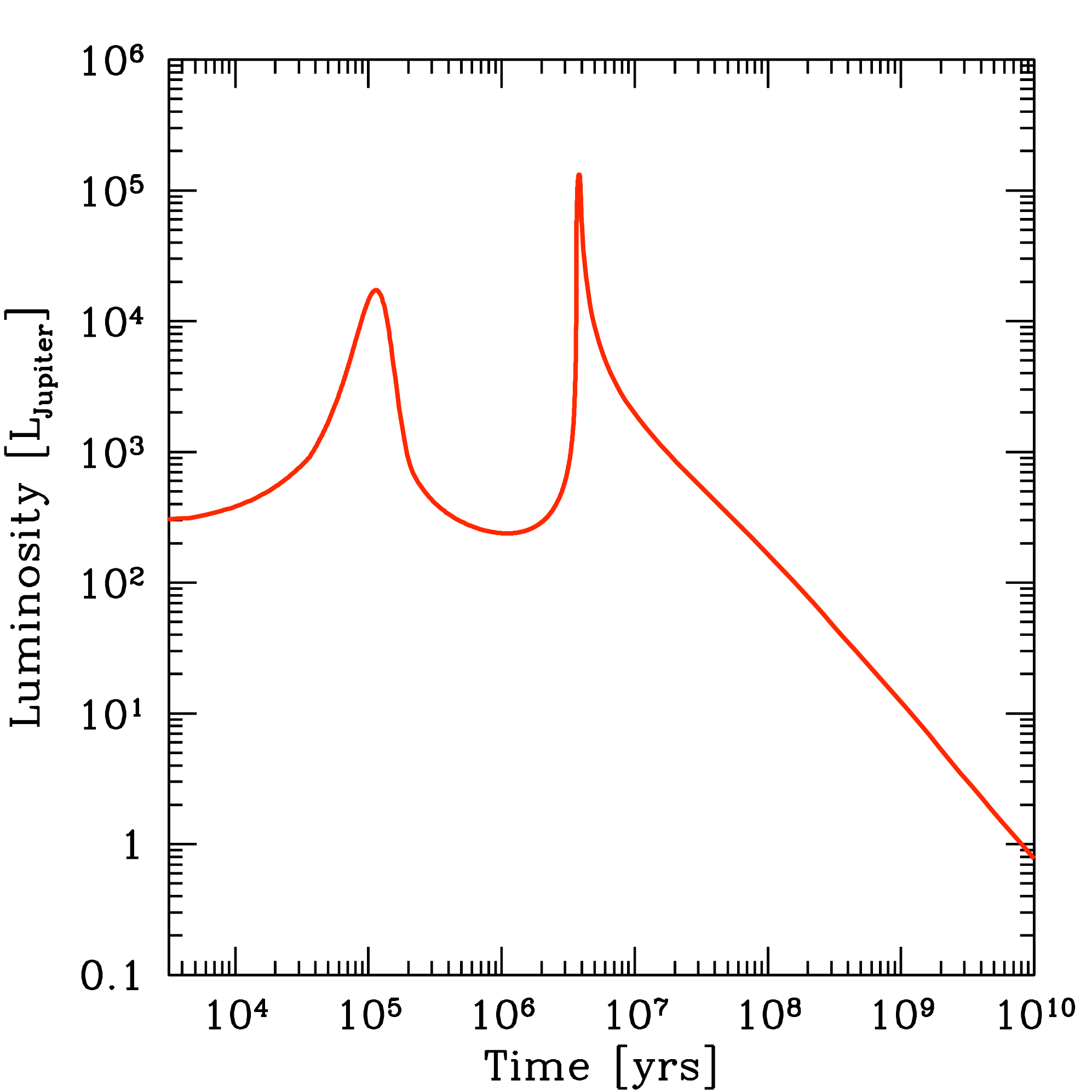}
     \end{minipage}
     \caption{Simulation for the in-situ formation of Jupiter. The top left panel shows the evolution of the core mass (red solid line), the envelope mass (green dotted line) and the total mass (blue solid line). The top right panels shows the accretion rate of solids (red solid line) and of gas (green dotted line). The limiting gas accretion rate is set to $10^{-3} \mearth$/yr. Note that the model is allowed to overshoot this value for a few time steps. The bottom left panel shows the evolution of the core radius (red solid line), the total radius (blue dashed line) and the capture radius (green dotted line). The latter radius is relevant for the capture of planetesimals. It is larger than the core radius due to the braking effect of the envelope. The outer radius is initially (during the attached regime) very large, as it is equal to the Hill sphere radius. At about 4 Myrs, it detaches from the nebula and collapses to a radius of initially about 2 Jupiter radii. Afterwards there is a slow contraction phase. The bottom right panel shows the internal luminosity of the planet. The first peak in the curve is due to the rapid accretion of the core (phase I), and the second to the runaway gas accretion/collapse phase.}\label{mordasinifig:coreaccretion} 
\end{figure*}

Figure \ref{mordasinifig:coreaccretion}  shows the formation and subsequent evolution of a Jupiter mass planet fixed at 5.2 AU for initial conditions equivalent to case J6 in \citet{pollackhubickyj1996} (initial solid surface density 10 g/cm$^{2}$, grain opacity in the envelope 2\% of nominal). In the calculation shown here the core density is variable, the luminosity is spatially constant in the envelope, but derived from total energy conservation \citep{papaloizounelson2005}. The limiting maximal gas accretion rate is simply set to $10^{-3} \mearth$/yr, and accretion is completely stopped once the total mass is equal to one Jupiter mass. In a full simulation \citep{alibertmordasini2005}, the maximal limiting accretion rate, as well as the termination of gas accretion is given by the decline of the gas flux in the disk caused by the evolution of the protoplanetary nebula.

The top left panel shows that three phases can be distinguished. In phase I, a solid core is built up. The solid accretion rate is large, as shown by the top right panel. The phase ends at when the planet has exhausted its feeding zone of planetesimals, which means that the planet reaches the isolation mass, which is of order 11.5 $\mearth$, in agreement with equation \ref{mordasinieq:miso}. In phase II, the accretion rates are low, and the planet must increase the feeding zone. This is achieved by the gradual accretion of an envelope: An increase in the gas mass leads to an increase of the  feeding zone of solids. Therefore the core can grow a little bit. This leads to a contraction of the external radius of the envelope. Gas from the disk streams in, leading to an increase of the envelope mass and so on. 

In phase III runaway gas accretion occurs. It starts at the crossover mass, i.e. when the core and envelope mass are equal (about 16.4 $\mearth$ in this simulation). At this stage, the radiative losses from the envelope can no more be compensated for by the accretional luminosity from the impacting planetesimals alone. The envelope has to contract, so that the new gas can stream in (note the quasi exponential increase of the gas accretion rate), which increases the radiative loss as the Kelvin Helmholtz timescale decreases strongly with mass in this regime, so that the process runs away, building up quickly a massive envelope. 

The existence of such a critical mass is intrinsic to the core-envelope setup and not dependent on detailed physics \citep{stevenson1982,wuchterl1993}. The critical core mass is typically of the order of 10-15 Earth masses, but it can be 1 to 40 $\mearth$ in extreme cases \citep{papaloizouterquem1999}. 

Shortly after the beginning of runaway gas accretion phase, the  limiting gas accretion rate is reached. The collapse phase starts which is actually a fast, but still hydrostatic contraction \citep{bodenheimerpollack1986,lissauerhubickyj2009} on a timescale of $\sim10^{5}$ years. The planet's surface detaches now from the surrounding nebula. The contraction continues quickly down to an outer radius of about 2 $R_{\rm Jupiter}$ (bottom left panel). 

Over the subsequent billion years, after the final mass is reached, slow contraction and cooling occurs. The different phases can also be well distinguished in the luminosity of the planet (bottom right curve), in particular the two maxima when a lot of gravitational binding energy is release during first the rapid accretion of the core, and then the runaway gas accretion and collapse phase.

The baseline formation model has many appealing features, producing a Jupiter like planet with an internal composition similar to what is inferred from internal structure model in a four times MMSN in a few million years. Note that in the calculation shown here it was assumed that all planetesimals can reach the central core. In reality, the shielding effect of a massive envelope prevents planetesimals of 100 km in radius as assumed here to reach directly the core for core masses larger than about six Earth masses \citep{alibertmousis2005}, and the planetesimals get instead dissolved in the envelope. 

The largest part of the evolution is spent in phase II. The length of this phase becomes uncomfortably close to protoplanetary disk lifetimes for lower initial solid surface densities, or for higher grain opacities in the envelope \citep{pollackhubickyj1996}. This is the so called timescale problem. Once migration is included, this problem can however be solved \citep{alibertmordasini2004}.

\section{From protoplanets to terrestrial planets}\label{mordasinisec:terrestrialplanets}
For terrestrial planet, the requirement that they form within the lifetime of the gaseous protoplanetary disk ($\leq10$ Myrs) can be dropped. Indeed we recall the final outcome of the planetesimal to protoplanets stage discussed in \ref{mordasinisec:planetesimalstoprotoplanets} in the inner part of the protoplanetary nebula (inside the iceline): A large number of oligarchs with masses $M\sim\miso$, i.e. between 0.01 to 0.1 $\mearth$.

Once the eccentricity damping caused by the gas disk (or by a sufficiently large population of small planetesimals) is gone, these oligarchs start to mutually pump up their eccentricities, and eventually the orbits of neighboring protoplanets cross, so that the final growth stage from $\miso$ to the final terrestrial planets with masses of order $\mearth$ starts. The system of big bodies evolves through a series of giant impacts to a state where the remaining planets have a configuration close to the smallest spacings allowed by long-term stability over Gyr timescales \citep{goldreichlithwick2004}.   

This long term stability manifests itself in the form of a sufficiently large mutual spacing of the bodies in terms of mutual Hill spheres, with typical final separation between the planets of a few ten $\rhill$ \citep{raymondbarnes2008}. Interestingly, such  basic architectures now become visible in the recently detected multi-planet extrasolar systems consisting of several low mass planets \citep{lovissegransan2010}. 

In the inner solar system, simulations (now based on direct N-body integration) addressing this growth stage must concurrently meet the following constraints \citep{raymondobrien2009}:  the observed orbits of the planets, in particular the small eccentricities (0.03 for the Earth);  the masses, in particular the small mass of Mars; the formation time of the Earth as deduced from isotope dating, about 50-100 Myrs; the bulk structure of the asteroid belt with a lack of big bodies; the relatively large water content of the Earth with a mass fraction of  $10^{-3}$ and last but not least, the influence from Jupiter and Saturn.

\begin{figure*}
       \includegraphics[width=\textwidth]{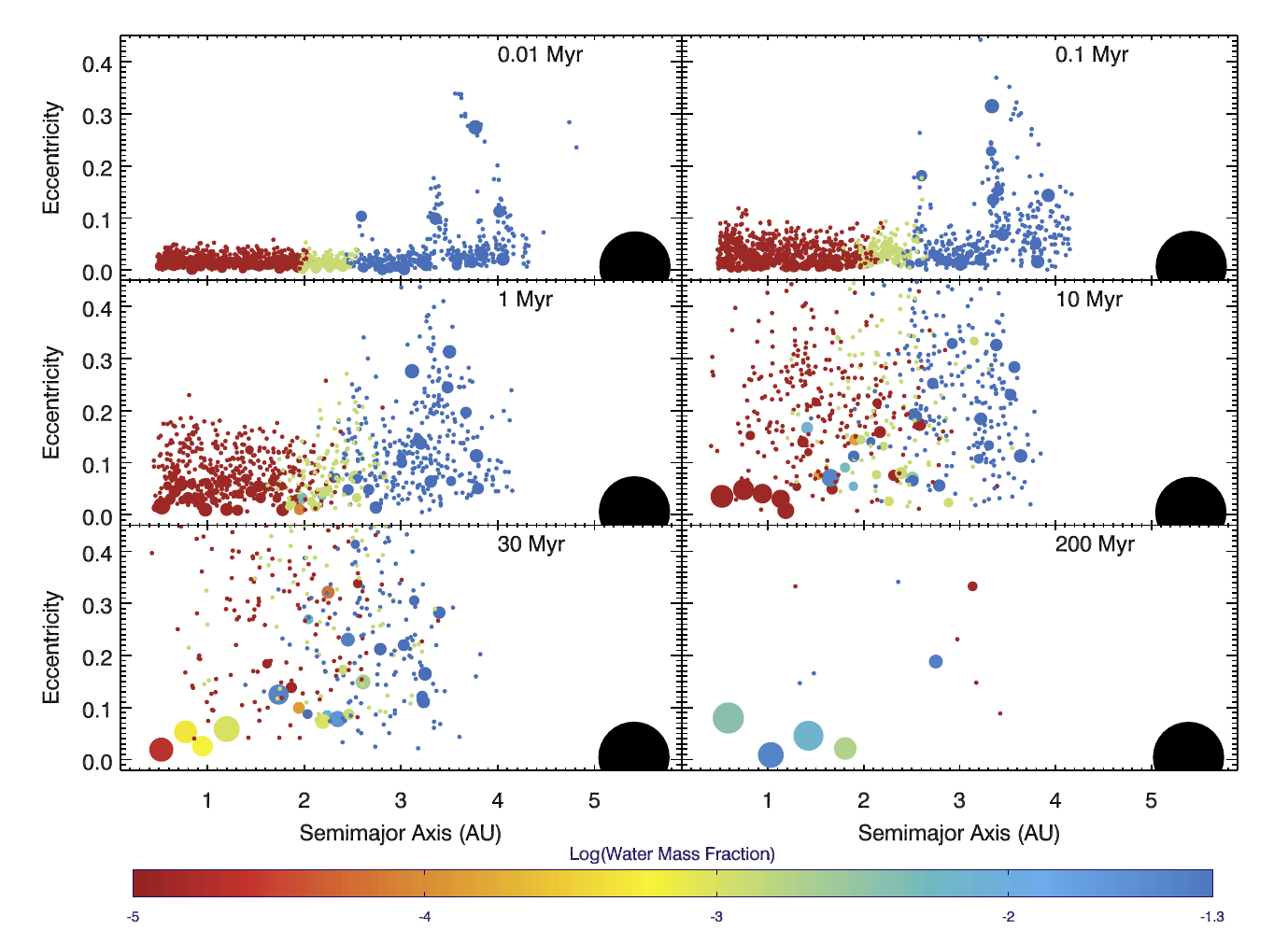}
     \caption{Six snapshots in time for an N-body simulation of  terrestrial planet formation by  \citet{raymondobrien2009}. The size of each body is proportional to its mass, while the color corresponds to the water content by mass, going from red (dry) to blue (5\% water). Jupiter is indicated as a large black circle while Saturn is not shown. Reproduced with permission from the author.}\label{mordasinifig:terrestrial}
 \end{figure*}

Figure \ref{mordasinifig:terrestrial} shows as an example six snapshots in time for the terrestrial planet formation in the inner solar system by \citet{raymondobrien2009}.  This simulation starts with roughly 100 0.01 to 0.1 $\mearth$ oligarchs, plus additional background planetesimals, as well as Jupiter and Saturn. One notes in the first panel the well defined eccentricity excitations at the places of mean motion resonances with the giant planets. In panel two and three, this resonant excitation spreads out.  During the stage of chaotic growth (until about 100 Myr), substantial radial mixing occurs, bringing water rich bodes in the inner system, as visible in the last three panels.  At the end, four terrestrial planets with masses between about 0.6 to 1.8 $\mearth$ have formed. The orbital distances, eccentricities, masses, formation timescales, and water content found in this simulation are approximatively in agreement with the actual solar system, but the Mars analogue is too massive, and there are three additional bodies in the asteroid belt. 

This simulation thus reproduces many important observed aspects, but not all of them. A main complication arises that the positions, eccentricities and masses of the giant planets at each moment in time are not exactly known, but significantly influence the formation of the terrestrial planets. 

\section{Orbital migration}\label{mordasinisec:migration}
Orbital migration occurs through the gravitational interaction of the planet with the protoplanetary disk. If the resulting torques exerted by the different parts of the disk onto the planet do not sum up to exactly zero, the planet will react on them by adjusting its angular momentum, i.e. its semimajor axis.

Migration comes in two different types: Low mass planets migrate in so called type I migration \citep{goldreichtremaine1980,tanakatakeuchi2002}, while massive planets which can open a gap in the gaseous disk migrate in type II \citep{linpapaloizou1986a}. Migration can be both a threat as a benefit for planet formation: On one hand, if migration happens on a timescale shorter than the growth timescale, planetary cores fall into the star before the can significantly grow \citep[e.g.][]{mordasinialibert2009b}. On the other hand, it allows planetary cores to grow beyond the isolation mass as cores always get into new regions of the disk where there are still fresh planetesimals to accrete, which reduces the formation time of giant planets, as the lengthy phase II (sect. \ref{mordasinisec:coreaccretion}) is skipped \citep{alibertmordasini2004}. 

\begin{figure*}
       \includegraphics[width=\textwidth]{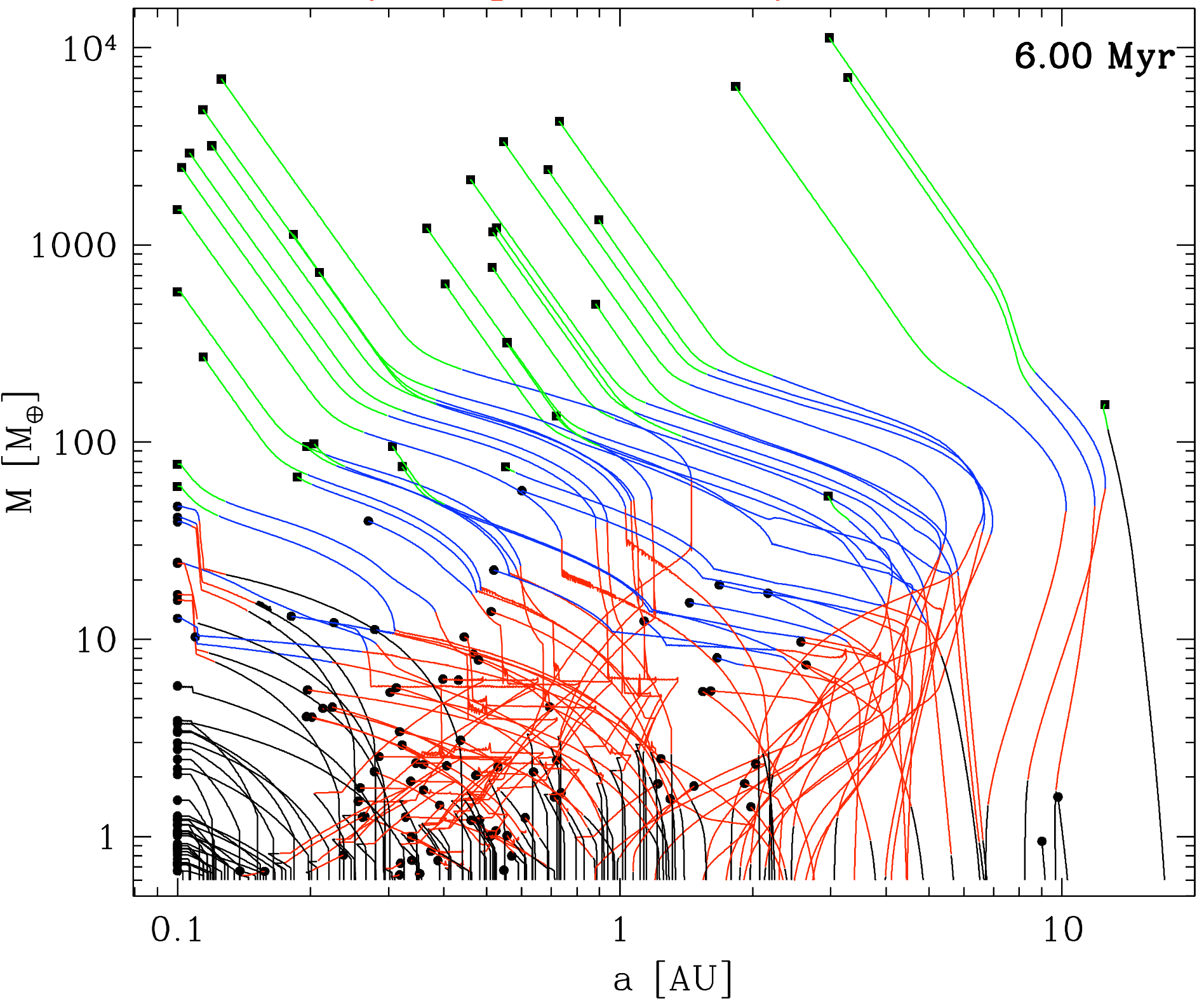}
     \caption{Exemples of planetary formation tracks in the mass distance plane. Each track corresponds to one embryo growing in a different protoplanetary disk. All seed embryos start with an initial mass of 0.6 $\mearth$, at different initial starting positions. Then they  grow and migrate towards their final position indicated by a large black symbol. The inner boarder of the computational domain is 0.1 AU. The colors correspond to different migration regimes \citep{dittkristmordasini2010}.}\label{mordasinifig:tracks} 
 \end{figure*}

Figure \ref{mordasinifig:tracks} \citep{dittkristmordasini2010} shows planetary formation tracks in the mass distance diagram. These tracks were calculated in an updated version of the population synthesis models \cite{mordasinialibert2009a} based on the extended core accretion model of \citet{alibertmordasini2005} which includes disk evolution and migration.  It shows how embryos with an initial mass of 0.6 $\mearth$ starting in different protoplanetary disks (characterized each by a disk mass, a solid content and a lifetime), and at different initial positions $\astart$, grow in mass and migrate. The final position of a planet is indicated by a large black symbol \citep[c.f.][]{mordasinialibert2009a}. 

In the plot, different colors indicate different migration regimes. Black stands for isothermal type I migration, red for unsaturated adiabatic type I, blue for saturated type I, and green finally for type II. These tracks were obtained using recent results on type I migration of \citet{paardekooperbaruteau2010,kleybitsch2009} and \citet{cridamorbidelli2007}. Depending on the local disk properties like the temperature and surface density gradients, both in- and outward migration can occur. The disk model is therefore of very high importance. It is also found that migration and accretion can strongly interact: Horizontal tracks for example occur when a planet migrates through a part of the disk which it has previously emptied from planetesimals while migrating outwards. On the other hand, gas runaway accretion and the associated mass growth can cause the switch to another migration regime. Note how Hot Jupiters form as the isolation mass limitation in the inner system can be overcome thanks to migration. 

\section{Selected questions}\label{mordasinisec:keyquestions}
The theory of planet formation which is linked to the discovery of extrasolar planets is one of the most active research fields of modern astronomy. In this final section we address two selected questions which are currently studied in the field.
 
\subsection{Luminosities of young planets}
A key property for several detection methods of extrasolar planets like direct imaging or interferometric methods is the luminosity of  (young) giant planets \cite[e.g.][]{absillebouquin2010}. As we have seen in section \ref{mordasinisec:coreaccretion}, is the luminosity a strongly function of time (and of planetary mass). The highest luminosities occur during the gas runaway accretion and collapse phase. The maximal values occurring then depend on the maximal possible gas accretion rate. For fast gas accretion one finds very high peak luminosities,  up to $\sim0.1 L_{\odot}$ for a Jupiter mass planet, but the phase of high luminosity is only very short, about $10^{4}$ years. Lower peak luminosities are found if the gas accretion rate is low, but the duration of the phase is correspondingly longer \citep{lissauerhubickyj2009}.  

This local energy input into the disk leads to the formation of a hot blob ($T\approx400-1500$ K) around the planet, with a size equal a few Hill spheres of the planet, corresponding to 0.1-1 AU for a growing giant planet a 5 AU. Such a feature should be detectable in the mid-IR, and might even cast shadows \citep{klahrkley2006,wolf2008}.

\begin{figure*}
\begin{minipage}{0.33\textwidth}
       \includegraphics[width=1.02\textwidth]{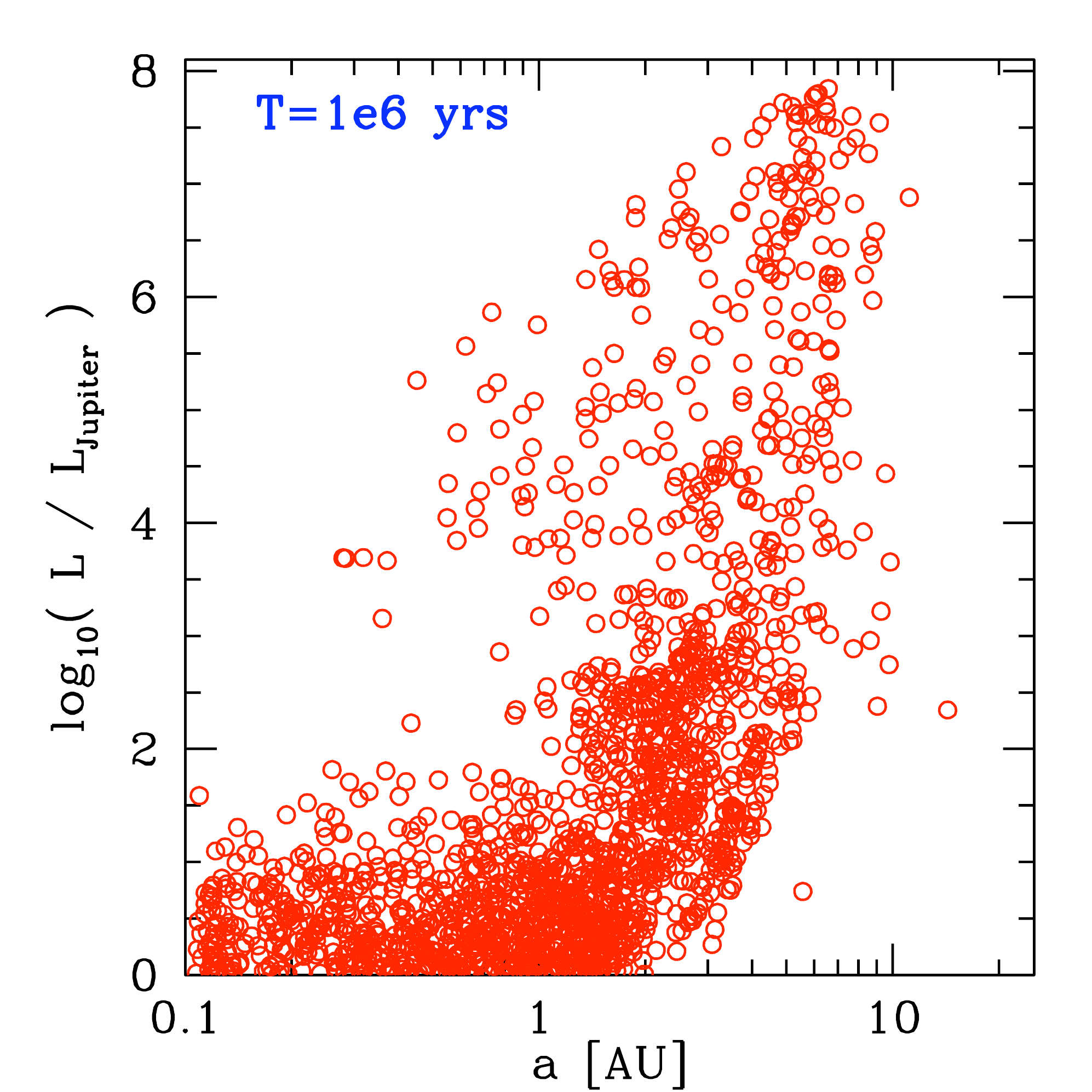}
     \end{minipage}\hfill
     \begin{minipage}{0.33\textwidth}
       \includegraphics[width=1.02\textwidth]{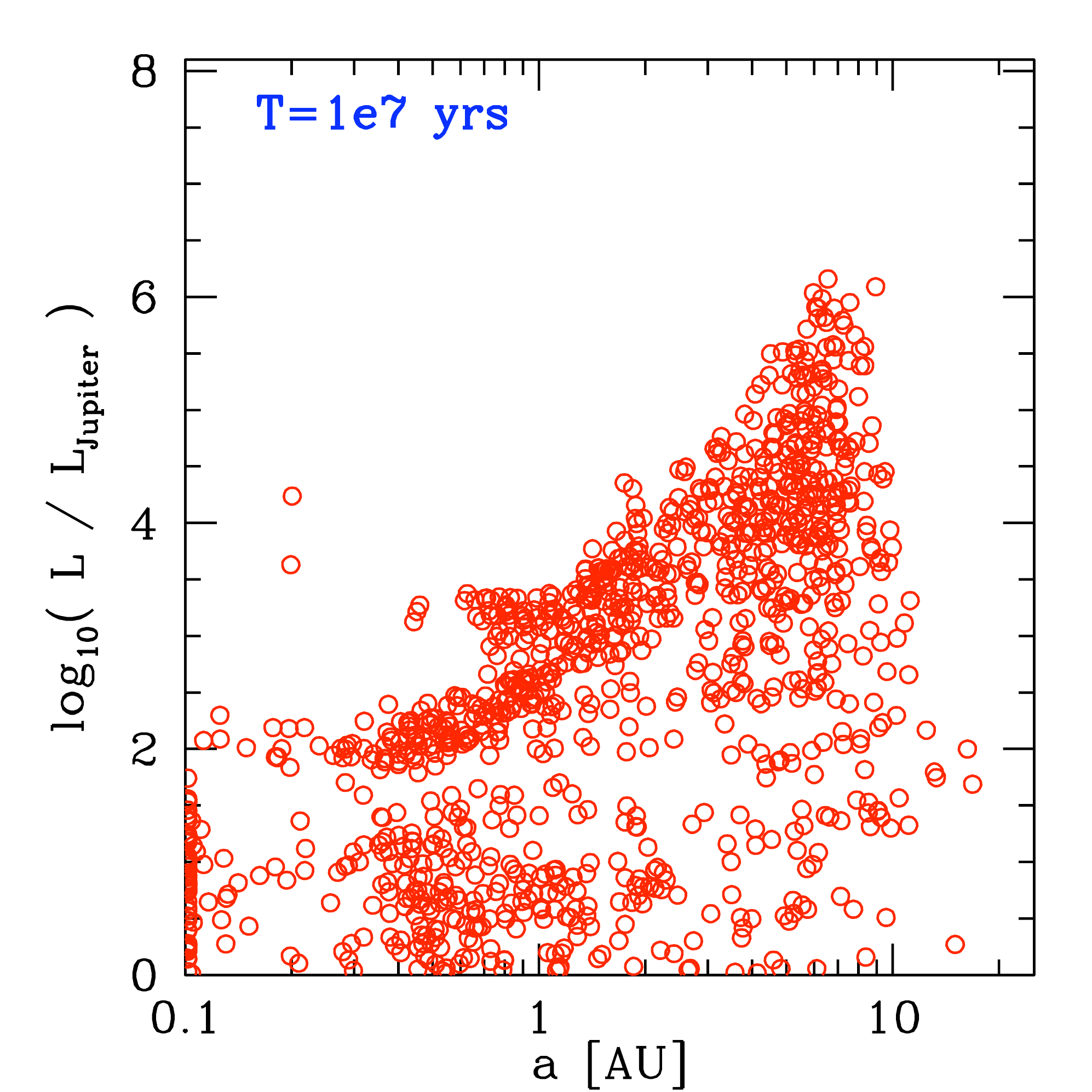}
     \end{minipage}\hfill
     \begin{minipage}{0.33\textwidth}
       \includegraphics[width=1.02\textwidth]{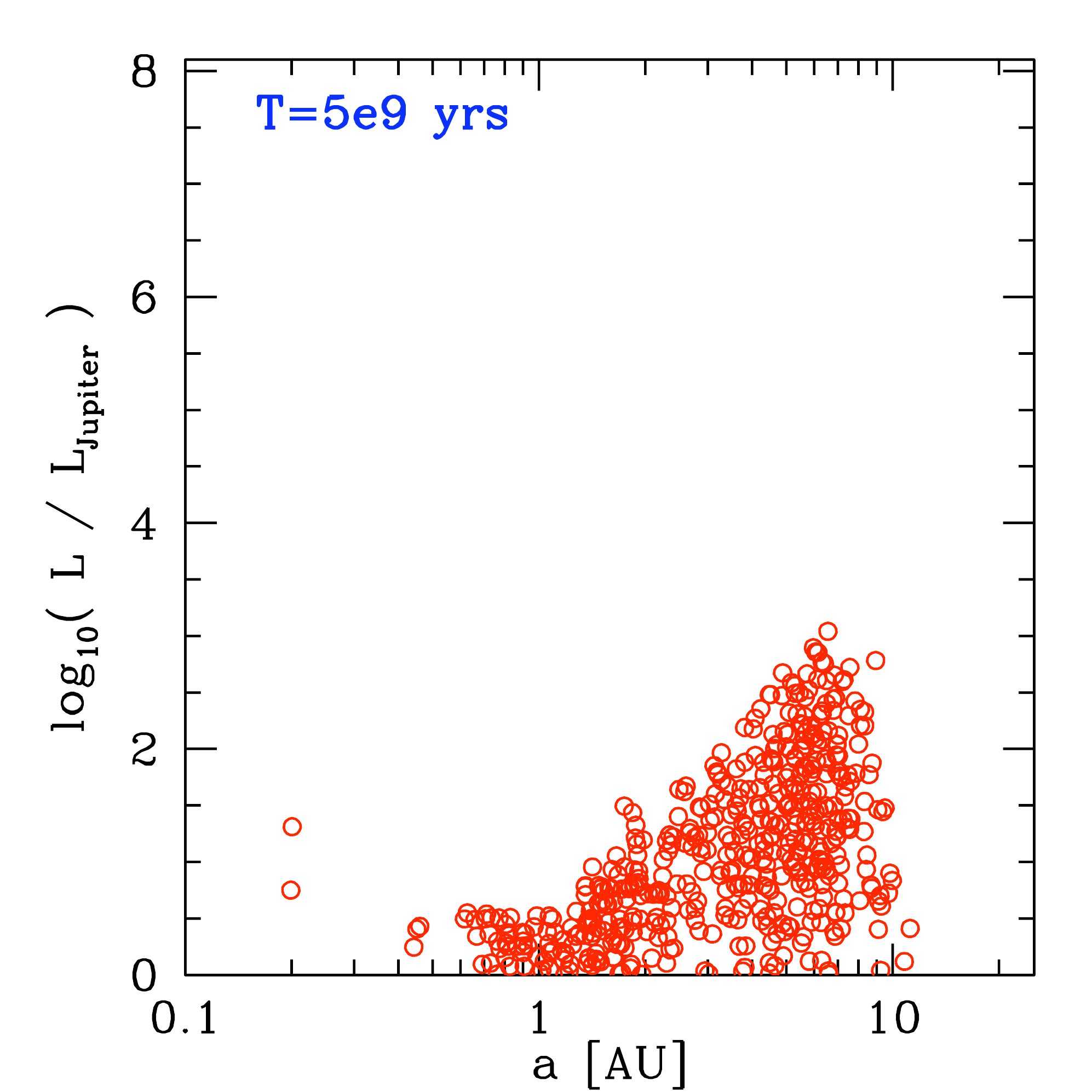}
     \end{minipage}
     \caption{Luminosity for a population of synthetic planets forming and evolving around a solar like star. The luminosity in units of Jupiter's luminosity today ($8.7\times10^{-10}L_{\odot}$) is shown as a function of distance, for three moments in time. The left panels is during the formation epoch, the middle panel soon after the protoplanetary disks have disappeared, and the right panels shows the situation after $5\times10^{9}$ years. Note that outward migration, or outward scattering is impossible in these simulation by construction, which likely leads to an underestimation of giant planets at large semimajor axes.  From Mordasini et al. in prep.}\label{mordasinifig:alumi} 
\end{figure*}

For observational surveys looking for planetary companions, it is relevant to know the distance-luminosity distribution as a function of time.  Figure \ref{mordasinifig:alumi} (Mordasini et al. in prep.) shows this for a synthetic population of planets forming around a solar like star. Note however that outward migration is not yet possible by construction in this set of models, which might be relevant in the context, as it lead to an underestimation of giant planets at large semimajor axes which are prime targets for such surveys. Note further that the luminosity of young giant planets is in general a topic which is still debated \citep{fortneymarley2005}.

The left panel shows the distance-luminosity plane at an age of the protoplanetary disks of $10^{6}$ years. One can see that there are some very bright planets with luminosities up to about 0.06 $L_{\odot}$. They are found somewhat outside the current position of Jupiter, at 6-7 AU. These are massive planets forming in solid rich disk, which get early in the gas runaway accretion/collapse phase, so that high gas accretion rates are possible. The much lower luminosities of the planets at smaller distances (inside about 1 AU) is caused in contrast by the accretion of planetesimals.

 The middle plot shows the situation at 10 million years, thus after all gaseous protoplanetary disks have disappeared. Gas accretion onto the planets has therefore ceased, but the planets are still quite luminous in this early phase of contraction. At the present age of the solar system, as shown in the right panel, luminosities have decreased by several orders of magnitude. The highest luminosities are caused by some $\sim30$ Jupiter mass objects. Two massive planets relatively close to the star are also visible.

\subsection{Where in the disk do giant planets form?}
The rate equations for solid accretion (sect. \ref{mordasinisec:rateequations}) indicate that a position not far outside the iceline is the sweet spot for giant planet formation within the core accretion paradigm, as massive cores can form still relatively quickly. This is confirmed by more complete simulations as can be seen in Fig. \ref{mordasinifig:astart} \citep{mordasinialibert2010}:

\begin{figure}\sidecaption
       \includegraphics[width=0.57\textwidth,height=7cm]{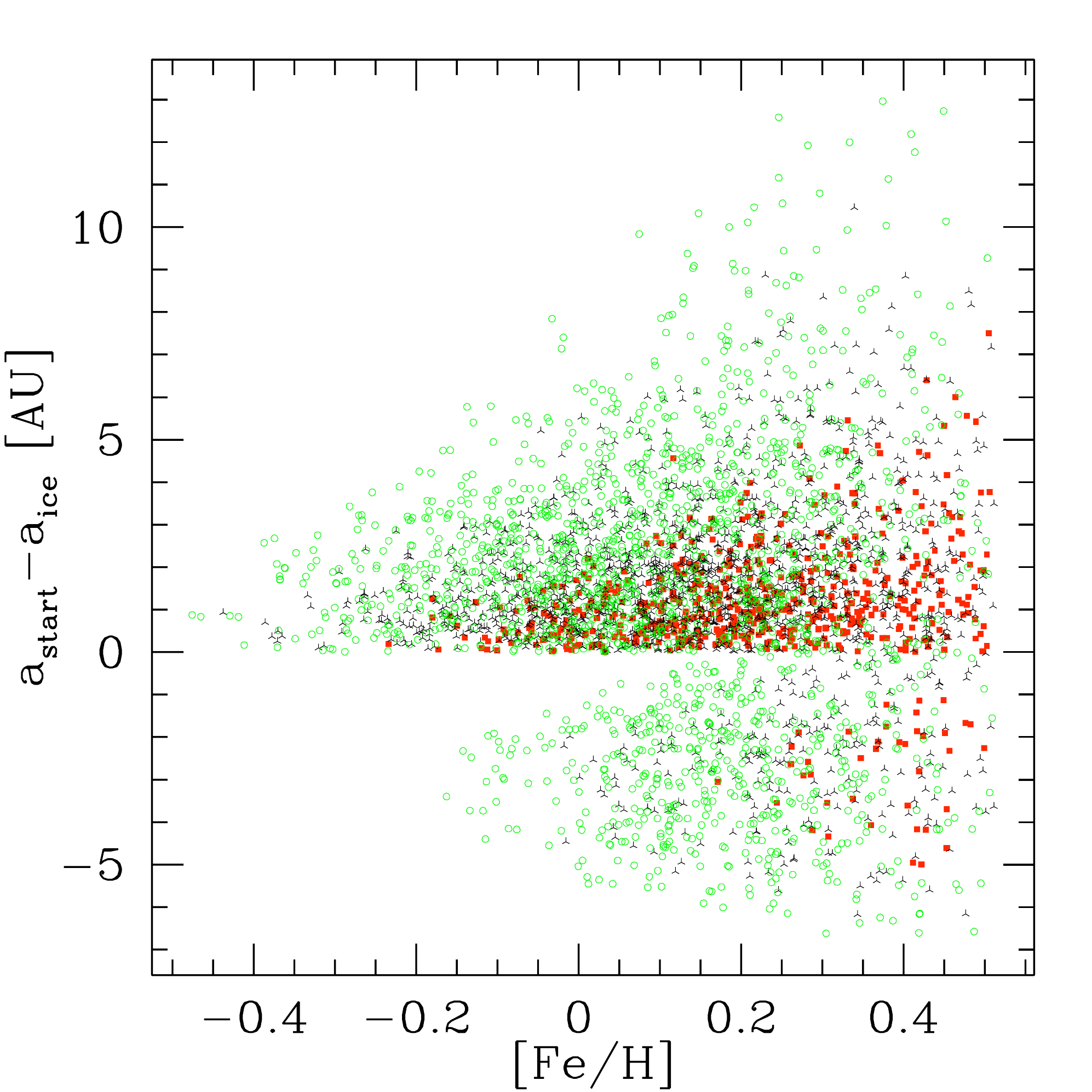}
     \caption{Distance $\astart$ from where planetary seed eventually becoming giant planets originate, relative to the location of the iceline $\aice$ in the corresponding protoplanetary disk. The distance is shown as a function of metallicity [Fe/H], i.e. the (logarithmic and normalized) dust to gas ratio in the disk. From \citet[][]{mordasinialibert2010}}\label{mordasinifig:astart}
 \end{figure}
 
The figure shows the initial location of planets $\astart$ that eventually grow more massive than 300 Earth masses (about 1 Jupiter mass), relative to the position of the iceline $\aice$ in their parent protoplanetary disk. The distance is shown as a function of [Fe/H], which measures the dust to gas ratio in the disk, normalized to the solar value ([Fe/H]=0). One sees that the typical position from where giant planet come is indeed a few AU outside the iceline, especially for low and medium [Fe/H] disks. At high [Fe/H], giant planets can form both inside, as well as clearly outside $\aice$ \citep[see also][]{idalin2004}. 

The location of the iceline $\aice$ itself is set, at least initially by the energy production due to viscous dissipation in the disk, and its radiation at the disk surface. Fits to the location of the iceline as a function of disk and stellar mass in this regime can be found in \citet{alibertmordasini2010}.  For a solar type star, $\aice$ lies between roughly 2 to 7 AU.  At later stages, when the disk becomes optically thin, the location of the iceline becomes determined by the energy input from the star \citep{idalin2004}, and lies at about 3 AU.
 
In the observed semimajor axis distribution of the extrasolar planets, there is an upturn in the frequency at a  semimajor axis of about 1 AU, which could be caused by this preferred starting position close to the iceline, and subsequent orbital migration \citep{mordasinialibert2009b,schlafumanlin2009}. This would thus be a direct imprint of disk properties on planetary properties. But one should keep in mind that the temperature and solid surface density structure in the disks  might be in reality much more complicated \citep{dzyurkevichflock2010} than assumed in the simple models used here. 

Observing planet formation directly  would therefore bring new, very valuable constraints for the theory of planet formation.

\begin{acknowledgements}
We thank Chris Ormel, Kai-Martin Dittkrist and Andreas Reufer for useful discussions. This work was supported in part by the Swiss National Science Foundation. Christoph Mordasini acknowledges the financial support as a fellow of the Alexander von Humboldt foundation. Yann Alibert is thankful for the  support by the European Research Council under grant 239605.
\end{acknowledgements}

\bibliographystyle{spbasic}      
\bibliography{mordasinilib}   

%
%

\end{document}